\documentclass[twocolumn,superscriptaddress,amsmath, amssymb, amsfonts,preprintnumbers,aps,prd,longbibliography,nofootinbib]{revtex4-1}

\usepackage{scalerel}
\usepackage{color}
\usepackage{amsmath,amsfonts,amssymb}
\usepackage[small,bf,hang]{caption}
\usepackage{slashed}
\usepackage{latexsym,epsfig}
\usepackage{dsfont}
\usepackage{extarrows}
\usepackage{hyperref}
\usepackage{array}

\def\moth{\mathsurround=0pt}
\newdimen\zo \zo=0pt

\def\tick{\leaders\hrule height 0.5ex depth 0pt \hskip 0.5pt}
\def\upboxfill{$\moth \setbox\zo\hbox{\tick}%
  \hskip 3pt\hbox to 0pt{$\tick$\hss}\hrulefill \hbox to 7.5pt{$\tick$\hss}$}

\def\dtick{\leaders\hrule height .34pt depth 0.5ex \hskip 0.5pt}
\def\downboxfill{$\moth \setbox\zo\hbox{\dtick}%
  \hskip 2pt\hbox to 0pt{$\dtick$\hss}\hrulefill \hbox to 2pt{$\dtick$\hss}$}

\def\bec{\begin{center}}
\def\ec{\end{center}}

\def\nn{\nonumber}

\def\be{\begin{equation}}
\def\ee{\end{equation}}
\def\bea{\begin{eqnarray}}
\def\eea{\end{eqnarray}}
\def\ba{\begin{array}}
\def\ea{\end{array}}

\begin{document}

\title{${\cal N}=1$ supersymmetry in $4D$ gravity with torsion and non-metricity}

\author{Cecilia Bejarano} 
\email{cbejarano@iafe.uba.ar}
\affiliation{Instituto de Astronom\'ia y F\'isica del Espacio (IAFE-UBA-CONICET), Ciudad Universitaria, Pabell\'on IAFE, Ciudad de Buenos Aires (1428), Argentina}

\author{Eric Lescano} 
\email{elescano@irb.hr}
\affiliation{Division of Theoretical Physics, Rudjer Bošković Institute, Bijenička 54, 10000 Zagreb, Croatia}

\begin{abstract}
We study the inclusion of fermionic degrees of freedom and ${\cal N}=1$ supersymmetry in 4-dimensional manifolds with arbitrary torsion and non-metricity tensors deforming the connection. We inspect the closure of local diffeomorphism, Lorentz and supersymmetric transformations on the vielbein and gravitino at leading order in fermions. The closure of a pair of supersymmetric transformations acting on the vielbein restricts the form of the spin connection. The dependency of the latter on the fundamental fields is exclusively given  by the vielbein, up to Lorentz and/or diffeomorphisms brackets, independently of the torsion or non-metricity deformations of the affine connection, which is not restricted by supersymmetry at this order. We have then studied the couple of the Rarita-Schwinger theory to a generic gravitational Lagrangian; the supersymmetric variation of the former restricts the form of the bosonic Lagrangian in order to accomplish invariance. Special cases where $R_{\mu \nu}{}^{ab}=0$, $Q_{\mu \nu \rho}=0$ or $T_{\mu \nu}{}^{\rho}=0$ are provided.

\end{abstract}

\maketitle

\section{Introduction}

In  Ref.~\cite{F}, Fock initiated the inclusion of fermionic degrees of freedom in General Relativity (GR). The standard procedure requires, on the one hand, fields with Grassmanian statistics and, on the other, a Clifford algebra for the gamma matrices. The fermionic fields are usually defined on the tangent space of a Riemannian manifold so the fundamental gravitational field is given by a tetrad (or vielbein). While the GR Lagrangian for the gravity sector is given by the Ricci scalar, fermionic contributions can be easily coupled to the Einstein-Hilbert Lagrangian depending on the spin of the Grassmanian field: a free spin-$\frac12$ field is coupled considering the Dirac Lagrangian while a free spin-$\frac32$ field is coupled using the Rarita-Schwinger (RS) Lagrangian \cite{RandS}. The latter is supersymmetric invariant meaning that if we couple the RS Lagrangian to a GR framework and we include supersymmetric transformations, the supersymmetric transformation of the RS Lagrangian compensates the transformation of the Ricci scalar. Furthermore in this scenario the Majorana spinor can be interpreted as a gravitino field \cite{WandZ}.

In the last years, several proposals were considered to deform the gravitational sector considering arbitrary torsion or non-metricity tensors. When the connection is given by the Levi-Civita connection (and the torsion and the non-metricity tensors vanish), the resulting theory is GR. For the curvature-less Weitzenbock connection the framework is known as Teleparallel Equivalent of GR, while for a gravitational theory defined by a torsion-free, flat connection we have Symmetric Teleparallel Equivalent of GR. All these theories share the same dynamics (known as geometrical trinity of gravity), but this is not true for the modified versions of the latter, i.e., $f(T)$-teleparallel theory \cite{f(T)1}-\cite{f(T)10}  and $f(Q)$-symmetric teleparallel theory \cite{f(Q)1}-\cite{f(Q)4}. In presence of spinors, these kind of proposals where studied in \cite{RecentSpin1}-\cite{RecentSpin3}. In vacuum solutions, both the torsion and the non-metricity are useful geometrical pieces to deform a connection with covariant extra terms, known as contortion and distortion. Particularly, when one includes fermionic matter, the commutator of a pair of covariant derivatives acting on the spinors contains new contributions. Moreover, in presence of a non-trivial non-metricity the covariant derivative of the gamma matrix is not null. 

The main goal of this work is to study how to couple the RS dynamics for backgrounds with arbitrary torsion and non-metricity, and to inspect the restrictions that ${\cal N}=1$ supersymmetry forces on the gravitational sector. Our main results are: On the one hand we study the closure of the symmetry transformations by  considering the leading order terms in fermions \cite{Tanii}. Interestingly enough, by taking into account arbitrary torsion and non-metricity tensors, the commutator of a pair of supersymmetric transformations on the vielbein forbids deformations on the spin-connection up to Lorentz and/or diffeomorphims bracket. This result is independent of the fermionic Lagrangian so it applies for spin-$\frac12$ fields as well as $\frac32$ fields. On the other hand, we study the restrictions on the gravitational theory when the RS Lagrangian is coupled. We performed the supersymmetric transformation of this Lagrangian in presence of arbitrary torsion and non-metricity tensors, and then we give the specific constraints for geometries with $R_{\mu \nu}{}^{ab}=0$, $Q_{\mu \nu \rho}=0$ or $T_{\mu \nu}{}^{\rho}=0$.  

The results of this work show agreement between two large fields of knowledge: on the one hand, the field of GR and its several modifications related to the inclusion of torsion and non-metricity as fundamental degrees of freedom and, on the other hand, the inclusion of fermionic matter related to the vacuum fields through supersymmetry. The current state of the art indicates several extension to this work, as we discussed in Section V.

\section{Field content, symmetries and covariant derivatives}\label{sec:fields}

We consider a fermionic matter field coupled to an arbitrary geometry. The fundamental fields are a four-dimensional Majorana spinor field, $\psi^{a}$, and a four-dimensional vielbein $e_{\mu}{}^{a}$ \footnote{We give our conventions in appendix \ref{Conventions}}. 

Majorana spinors satisfy a reality condition,
\bea
\psi_{a}^c = C \bar \psi^{T} = \psi_{a} \, ,
\eea
where $C$ is a charged conjugation matrix which satisfies,
\bea
C^{-1} \gamma^{a} C = - \gamma^{a T} \, , \quad C^{T} = - C \, .
\eea 
The gamma matrices fulfill a Clifford algebra 
\bea
\Big\{ \gamma^{a},\gamma^{b} \Big\} = 2 \eta^{a b} \, ,
\label{Calgebra}
\eea
with $\eta^{a b}$ the inverse of the Minwkowski metric $\eta_{a b}$.
In turn, the metric tensor is obtained from the orthonormality condition as
\bea
e_{\mu}{}^{a} \eta_{a b} e_{\nu}{}^{b} = g_{\mu \nu} \, .
\eea

The symmetry rules of the fundamental fields are given by
\bea
\label{Symmetry1}
\delta e_{\mu}{}^{a} & = & \xi^{\nu} \partial_{\nu} e_{\mu}{}^{a} + \partial_{\mu} \xi^{\nu} e_{\nu}{}^{a} + e_{\mu}{}^{b} \Lambda_{b}{}^{a} + \frac12 \bar \epsilon \gamma^{a} \psi_{\mu} \\
\delta \psi^{a} & = & \xi^{\nu} \partial_{\nu} \psi^{a} + \psi^{b} \Lambda_{b}{}^{a} - \frac14 \Lambda_{bc} \gamma^{bc} \psi^{a} + e^{\mu}{}_{a} \nabla_{\mu} \epsilon \, ,
\label{Symmetry2}
\eea
where $\xi^{\mu}$, $\Lambda_{a}{}^{b}$ and $\epsilon$ are arbitrary parameters for local diffeomorphisms, Lorentz and supersymmetry transformations, respectively. From Eq.~(\ref{Symmetry1}) one can observe that the vielbein transforms as a vector with respect to infinitesimal diffeomorphims and local Lorentz transformations, while the spinor field $\psi^{a}$ is a scalar under infinitesimal diffeomorphims but transforms as a vector and a spinor with respect to local Lorentz transformations.

The covariant derivative on these fields is expressed in terms of the affine connection, $\Gamma_{\mu \nu}^{\rho}$, and the spin connection, $\omega_{\mu}{}^{a}{}_{b}$, i.e., 
\bea
\nabla_{\mu} e_{\nu}{}^{a} & = & \partial_{\mu} e_{\nu}{}^{a} - \Gamma_{\mu \nu}^{\sigma} e_{\sigma}{}^{a} + \omega_{\mu}{}^{a}{}_{b} e_{\nu}{}^{b} \, , \\
\nabla_{\mu} \psi_{a} & = & \partial_{\mu} \psi_{a} + \omega_{\mu a}{}^{b} \psi_{b} + \frac{1}{4} \omega_{\mu b c} \gamma^{bc} \psi_{a} \, .
\eea

We decompose the affine connection in the usual way,
\bea
\Gamma_{\mu \nu}{}^{\rho} = \tilde \Gamma_{\mu \nu}{}^{\rho} - K_{\mu \nu}{}^{\rho} - N_{\mu \nu}{}^{\rho} \,
\label{affine}
\eea
where $\tilde \Gamma_{\mu \nu}{}^{\rho}$ is the Levi-Civita connection and
\bea
K_{\mu \nu \rho} & = & \frac12 (-T_{\mu \nu \rho} + T_{\nu \rho \mu} - T_{\rho \mu \nu}) \nn \\
N_{\mu \nu \rho} & = & \frac12 (Q_{\mu \nu \rho} + Q_{\nu \mu \rho} - Q_{\rho \mu \nu})
\eea
the contorsion and the distorsion, respectively,
with the non-metricity and the torsion defined as follows
\bea
\label{nonm}
\nabla_{\mu} g_{\nu \rho}& = & Q_{\mu \nu \rho} \, , \\ T_{\mu \nu}{}^{\rho} & = & \Gamma_{\mu \nu}^{\rho} - \Gamma_{\nu \mu}^{\rho} \, .
\eea
In this work we will focus on the leading order in fermionic contributions, so $\Gamma$ is a generic affine connection and we do not need to impose a particular fermionic torsion, which is a requirement in the full picture \cite{Tanii}. Using the previous definitions, we impose the following vielbein compatibility in presence of torsion and non-metricity,
\bea
\nabla_{\mu} e_{\nu}{}^{a} = N_{\mu \nu}{}^{\rho} e_{\rho}{}^{a} + K_{\mu \nu}{}^{\rho} e_{\rho}{}^{a} \, ,
\label{ecov}
\eea
or, equivalently,
\bea
\tilde \nabla_{\mu} e_{\nu}{}^{a} \equiv \partial_{\mu} e_{\nu}{}^{a} - \tilde \Gamma_{\mu \nu}^{\sigma} e_{\sigma}{}^{a} + \omega_{\mu}{}^{a}{}_{b} e_{\nu}{}^{b} = 0 \, .
\label{compatibility}
\eea
The previous expression implies that the spin connection is defined as in GR or, in other words, it does not contain extra tensorial deformations related to the torsion and/or the non-metricity. Moreover, from Eq.~(\ref{ecov}) it is straightforward to obtain
\bea
2 \nabla_{\rho} e_{(\mu}{}^{a} \eta_{ab} e_{\nu)}{}^{b} = Q_{\rho \mu \nu} \, ,
\eea
in agreement with Eq.~(\ref{nonm}).

We can now define the curved version of the Majorana spinor field, $\psi^{\mu}=e^{\mu}{}_{a} \psi^{a}$ and the curved version of the gamma matrix, $\gamma^{\mu}=e^{\mu}{}_{a} \gamma^a$ which satisfies
\bea
\Big\{ \gamma^{\mu},\gamma^{\nu} \Big\} = 2 g^{\mu \nu} \, . 
\eea
The covariant derivative acting on a curved gamma matrix is given by
\bea
\nabla_{\mu} \gamma^{\nu} = \frac12 Q_{\mu}{}^{\nu}{}_{\rho} \gamma^{\rho} \, ,
\eea
and the commutator of a pair of covariant derivatives acting on a generic spinor reads
\bea
\Big[\nabla_{\mu},\nabla_{\nu}\Big] \epsilon = && \frac14 R_{\mu \nu}{}^{ab}(\omega) \gamma_{ab} \epsilon - 2 \Gamma_{[\mu \nu]}^{\rho} \nabla_{\rho} \epsilon \nn \\ && - \frac12 \omega_{\mu}{}^{ab} \nabla_{\nu} \gamma_{ab} \epsilon
\label{rel1}
\eea
while the same communtator acting on a the spinor field gives
\bea
\Big[\nabla_{\mu},\nabla_{\nu}\Big] \psi_{\rho} = && \frac14 R_{\mu \nu}{}^{ab}(\omega) \gamma_{ab} \psi_{\rho} - R_{\mu \nu}{}^{\sigma}{}_{\rho}(\Gamma) \psi_{\sigma} \nn \\ &&
- 2 \Gamma_{[\mu \nu]}^{\sigma} \nabla_{\sigma} \psi_{\rho} - \frac12 \omega_{\mu}{}^{ab} \nabla_{\nu} \gamma_{ab} \psi_{\rho} \label{rel2}
\, .
\eea
In the next part of the work we compute the closure of the different transformations and, particularly, we analyze the role of Eq.~(\ref{compatibility}) in the closure of the supersymmetric transformations.


\section{Closure of the transformations}\label{sec:closure}
The closure of a pair of symmetry transformations is satisfied when
\bea
\Big[\delta_{1},\delta_{2} \Big]\{e_{\mu}{}^{a},\psi^a\} = \delta_{3} \{e_{\mu}{}^{a},\psi^a\}\, .
\eea
The closures are given by the following brackets, 
\bea
\Big[ \delta_{\xi_{1}}, \delta_{\xi_{2}} \Big] & = & \delta_{\xi_{3}} \, , \quad \xi^{\mu}_{3}= 2 \xi^{\nu}_{[2} \partial_{\nu} \xi_{1]}^{\mu} \, . \\
\Big[ \delta_{\Lambda}, \delta_{\xi} \Big] & = & \delta_{\tilde \Lambda} \, , \quad \tilde \Lambda_{ab} = \xi^{\mu} \partial_{\mu}\Lambda_{ab} \, . \\
\Big[ \delta_{\epsilon}, \delta_{\xi} \Big] & = & \delta_{\tilde \epsilon} \, , \quad \tilde \epsilon = \xi^{\mu} \partial_{\mu} \tilde \epsilon \, . \\
\Big[ \delta_{\Lambda_{1}}, \delta_{\Lambda_{2}} \Big] & = & \delta_{\Lambda_{3}} \, , \quad \Lambda_{3c}{}^{a} = 2 \Lambda_{[1 c}{}^{b} \Lambda_{2]b}{}^{a} \, . \\
\Big[ \delta_{\Lambda}, \delta_{\epsilon} \Big] & = & \delta_{\tilde \epsilon} \, , \quad \tilde \epsilon = \frac14 \Lambda_{ab} \gamma^{ab} \epsilon \, ,
\eea
where $\xi, \epsilon, \Lambda$ are the infinitesimal parameters of the supersymmetric, diffeomorphism, Lorentz transformations, respectively.
It is worth to note that the previous closures are satisfied in any gravity theory. However, the closure $[\delta_{\epsilon_{1}}, \delta_{\epsilon_{2}}]$ forces Eq.~(\ref{compatibility}) up to a Lorentz/diffeomorphism bracket. Let us illustrate this point performing the explicit computation,
\bea
\Big[ \delta_{\epsilon_{1}}, \delta_{\epsilon_{2}} \Big] e_{\mu}{}^{a} = \frac12 \bar \epsilon_{2} \gamma^{a} \nabla_{\mu} \epsilon_{1} - (1 \leftrightarrow 2)  \, .
\label{eq1}
\eea
Using the explicit form of the covariant derivative on the spinor $\epsilon_{1}$, the Clifford algebra and the following properties
\bea
\bar \epsilon_{2} \gamma^{a} \partial_{\mu} \epsilon_{1} & = & - \partial_{\mu} \bar{\epsilon}_{1} \gamma^{a} \epsilon_{2} \, , \nn \\ \bar \epsilon_{2} \gamma^{abc} \epsilon_{2} & = & \bar{\epsilon}_{1} \gamma^{abc} \bar \epsilon_{2} \, , \nn
\eea
the expression (\ref{eq1}) can be written as
\bea
\Big[ \delta_{\epsilon_{1}}, \delta_{\epsilon_{2}} \Big] e_{\mu}{}^{a} = \frac12 \partial_{\mu}(\bar \epsilon_{2} \gamma^{a}\epsilon_{1}) + \frac12 \bar \epsilon_{2} \gamma^{c} \omega_{\mu}{}^{a}{}_{c} \epsilon_{1} \, .
\eea
The previous terms can be written as an infinitesimal diffeomorphism transformation with $\tilde \xi^{\nu}= \frac12 e^{\nu}{}_{a} \bar \epsilon_{2} \gamma^{a} \epsilon_{1}$,
\bea
\delta_{\tilde \xi} e_{\mu}{}^{a} & = & e^{\nu}{}_{b} \bar \epsilon_{2} \gamma^{b} \epsilon_{1} \partial_{[\nu} e_{\mu]}{}^{a} + \frac12 \partial_{\mu}(\bar \epsilon_{2} \gamma^{a} \epsilon_{1}) \, \nn \\ & = &
- \frac12 e^{\nu}{}_{b} \bar \epsilon_{2} \gamma^{b} \epsilon_{1} \omega_{\nu}{}^{a}{}_{c} e_{\mu}{}^{c} \nn \\ && + \frac12 \bar \epsilon_{2} \gamma^{c} \epsilon_{1} \omega_{\mu}{}^{a}{}_{c} e_{\mu}{}^{c} + \frac12 \partial_{\mu}(\bar \epsilon_{2} \gamma^{a} \epsilon_{1}) \, , 
\eea 
where in the last step we use the vielbein compatibility given by Eq.~(\ref{compatibility}). Finally,
\bea
\Big[ \delta_{\epsilon_{1}}, \delta_{\epsilon_{2}} \Big] = \delta_{\tilde \xi} + \delta_{\tilde \Lambda} \, , \quad \tilde \Lambda^{a}{}_{c} = \frac12 e^{\nu}{}_{b} \bar \epsilon_{2} \gamma^{b} \epsilon_{1} \omega_{\nu}{}^{a}{}_{c} \, .
\eea
Then, the closure of the supersymmetry algebra forces the use of the standard spin connection before introducing dynamics. This means that the affine connection can be generic, but the vielbein compatibility cannot contain terms with torsion or non-metricity. 

In the next part of the work we explore the RS Lagrangian. We explicitly show the conditions which arise from supersymmetric invariance in presence of a non-trivial torsion and non-metricity. We will focus on leading order terms in fermions, and that is why we omit the study of $\Big[ \delta_{\epsilon_{1}}, \delta_{\epsilon_{2}} \Big] \psi^{a}$ in the present section.


\section{Coupling the Rarita-Schwinger Lagrangian}\label{sec:lagrangian}
The minimal matter Lagrangian that can be coupled to a gravity theory to turn the latter into a supergravity theory (with ${\cal N}=1$ supersymmetric invariance) is the RS Lagrangian which can be understood as the generalization of the Dirac Lagrangian for spin-$\frac32$ fields. We then consider the following action principle,
\bea
S & = & \int d^4x ({\cal L}_{\textrm{gravity}} + {\cal L}_{RS}) \, ,
\eea
with ${\cal L}_{\textrm{gravity}}$ a generic gravitational vacuum Lagrangian for gravity and
\bea
{\cal L}_{RS} & = & - e \bar \psi_{\mu} \gamma^{\mu \nu \rho} \nabla_{\nu} \psi_{\rho} + \mathcal{O}(\psi^4) \, .
\label{RSLag}
\eea

Our goal now is to compute the supersymmetric variation of the matter contributions in order to find restrictions on the gravity sector. For this reason we apply the last transformations of equations (\ref{Symmetry1}) and (\ref{Symmetry2}) on Eq.~(\ref{RSLag}). After a lengthy but straightforward computation the supersymmetric variation of the RS Lagrangian reads
\bea
\delta_{\epsilon} {\cal L}_{RS} = && - \frac12 e \bar \epsilon Q_{\mu}{}^{\sigma}{}_{\sigma} \gamma^{\mu \nu \rho} \nabla_{\nu} \psi_{\rho} + e \bar \epsilon \nabla_{\mu}\gamma^{\mu \nu \rho} \nabla_{\nu} \psi_{\rho} \nn \\ && + e \bar \epsilon \gamma^{\mu \nu \rho} \nabla_{\mu} \nabla_{\nu} \psi_{\rho} - e \bar \psi_{\mu} \gamma^{\mu \nu \rho} \nabla_{\nu} \nabla_{\rho} \epsilon \, .
\label{fervariation}
\eea
We provide in Appendix \ref{Gamma} the products between gamma matrices relations required for this computation. The new terms in this expression are related to non-metricity effects (first line) and to torsion contributions (commutators of the covariant derivatives in second line) through equations (\ref{rel1}) and (\ref{rel2}). 

From the supersymmetric variation of Eq.~(\ref{fervariation}) one learns lessons for different proposals of theories of gravity when $R_{\mu \nu}{}^{ab}$, $Q_{\mu \nu \rho}$ or $T_{\mu \nu}{}^{\rho}$ vanish. We summarize our findings in Table \ref{tabcon}.

\begin{widetext}
\begin{center}
\begin{table}
 \begin{tabular}{|c|c|c|}
 \hline
 $e f(Q,T)$ & $R_{\mu \nu}{}^{ab}=0$ & 
 $\begin{aligned} \delta_{\epsilon} f(Q,T) & = \frac12 e \bar \epsilon Q_{\mu}{}^{\sigma}{}_{\sigma} \gamma^{\mu \nu \rho} \nabla_{\nu} \psi_{\rho} - e \bar \epsilon \nabla_{\mu}\gamma^{\mu \nu \rho} \nabla_{\nu} \psi_{\rho}\\ 
 & + \frac{1}{2} e \bar \epsilon \gamma^{\mu \nu \rho} (2 \Gamma_{[\mu \nu]}^{\sigma} \nabla_{\sigma}\psi_{\rho} + \frac12 \omega_{\mu ab} \nabla_{\nu} \gamma^{ab} \psi_{\rho})\\ 
 & - \frac{1}{2} e \bar \psi_{\mu} \gamma^{\mu \nu \rho} (2 \Gamma_{[ \nu \rho]}^{\sigma} \nabla_{\sigma}\epsilon + \frac12 \omega_{\nu ab} \nabla_{\rho} \gamma^{ab} \epsilon) 
\end{aligned}$\\ 
 \hline
 $e e^{\mu}{}_{a} e^{\nu}{}_{b} R_{\mu \nu}{}^{ab} + f(T)$ & $Q_{\mu \nu \rho} = 0$ & $\begin{aligned}\delta_{\epsilon} f(T) = - \frac12 e \bar \epsilon \gamma^{\mu \nu \rho} T_{\nu \rho}{}^{\sigma} \nabla_{\sigma} \psi_{\rho} + \frac12 e \bar \psi_{\mu} \gamma^{\mu \nu \rho} T_{\nu \rho}{}^{\sigma} \nabla_{\sigma} \epsilon 
 \end{aligned}$ \\
 \hline
 $e e^{\mu}{}_{a} e^{\nu}{}_{b} R_{\mu \nu}{}^{ab} + f(Q)$ & $T_{\mu \nu}{}^{\rho}=0$ & 
 $\begin{aligned} \delta_{\epsilon} f(Q) & = \frac12 e \bar \epsilon Q_{\mu}{}^{\sigma}{}_{\sigma} \gamma^{\mu \nu \rho} \nabla_{\nu} \psi_{\rho} - e \bar \epsilon \nabla_{\mu}\gamma^{\mu \nu \rho} \nabla_{\nu} \psi_{\rho} \\ & - \frac{1}{2} e \bar \epsilon \gamma^{\mu \nu \rho} ( Q_{\mu \nu}{}^{\sigma} \nabla_{\sigma}\psi_{\rho} - \frac12 \omega_{\mu ab} \nabla_{\nu} \gamma^{ab} \psi_{\rho}) \\ & + \frac{1}{2} e \bar \psi_{\mu} \gamma^{\mu \nu \rho} ( Q_{\nu \rho}{}^{\sigma} \nabla_{\sigma}\epsilon - \frac12 \omega_{\nu ab} \nabla_{\rho} \gamma^{ab} \epsilon)
 \end{aligned}$ \\ 
 \hline
 \end{tabular}
\caption{\label{tabcon} The first two columns correspond to particular bosonic Lagrangians which the Rarita-Schwinger Lagrangian can be coupled to. In the third column we show the constraints imposed by  ${\cal N}=1$ supersymmetric invariance.}
\end{table}
\end{center}
\end{widetext}

The last two rows of the table reduce to the well known Einstein-Hilbert Lagrangian when $T_{\mu \nu}{}^{\rho}=0$ or
$Q_{\mu \nu \rho}=0$, respectively. In that case (i.e., GR), the full action is ${\cal N}=1$ supersymmetric invariant due to
\bea
\delta_{\epsilon} (e e^{\mu}{}_{a} e^{\nu}{}_{b} R_{\mu \nu}{}^{ab}) = && \frac12 e \bar \epsilon \gamma_{\rho} \psi^{\rho}  e_{a}{}^{\mu} e_{b}{}^{\nu} R_{\mu \nu}{}^{ab}(\omega) \nn \\ && - e \bar \epsilon \gamma^{\mu} \psi_{a} e_{b}{}^{\nu} R_{\mu \nu}{}^{ab}(\omega) \, .
\eea
The previous terms cancel the terms of Eq.~(\ref{fervariation}) which are independent of $T_{\mu \nu}{}^{\rho}$ and $Q_{\mu \nu \rho}$.


\section{Conclusions and future directions}

We have analyzed the inclusion of ${\cal N}=1$ supersymmetric transformations in a general gravity framework, where the torsion and the non-metricity tensors are arbitrary. We have studied the closure of local Lorentz, diffeomorphisms and ${\cal N}=1$ supersymmetry transformations on the fundamental fields at leading order in fermions (bilineal contributions at the level of the action). We have found that the closure of a pair of supersymmetric transformations acting on the vielbein reproduce a diffeomorphism and a Lorentz transformation when the spin connection satisfies Eq.~(\ref{compatibility}), and all the possible deformations of the spin connection are restricted by the closure of the supersymmetric transformations, although they do not restrict the deformations of the affine connection (Eq.~(\ref{affine})). The inspection of the closure of the symmetry transformations is independent to the dynamics. 

On the other hand, we have obtained the form of the supersymmetric transformation in Eq.~(\ref{fervariation}) by taking into account arbitrary non-metricity and torsion tensors. In this way, the gravitational Lagrangian results constrained in order to accomplish supersymmetric invariance. In the present work, we have shown the explicit variations of ${\cal L_{\textrm{gravity}}}$ for three particular cases where $R_{\mu \nu}{}^{ab}$, $Q_{\mu \nu \rho}$ or $T_{\mu \nu}{}^{\rho}$ vanish. 

We close this section with some possible follow-up projects:
\begin{itemize}
\item The fermionic contributions at the level of the action principle might be used to establish new equivalences between gravity theories. The equation of motion for the fermionic field(s) might be useful to cure gravitational inequivalences considering different bilinears contributions, even for Lagrangians that are not supersymmetric.

\item Different constructions of $f(Q)$ cosmologies \cite{CosmoQ1}-\cite{CosmoQ6} requiere non-linear $f(Q)$ Lagrangians to include dark components of the Universe from a variational principle. The inclusion of ${\cal N}=1$ supersymmetry  in $f(Q)$ cosmological scenarios, which can be easily related to string cosmology scenarios \cite{Gasperini}, allows the possibility of condensates in these models. This, in turn, could be useful to explain the statistically-significant tension in the realistic cosmological scenarios \cite{CosmoR}. 

\item It would be useful to extend the notion of supersymmetric modified gravities into the language of T-duality invariant cosmological frameworks \cite{T-duality1}-\cite{T-duality5}, where the notion of generalized geometry arises. The agreement between Riemannian geometries and the generalized geometry required to manifest T-duality invariance was studied in \cite{Victor} and more recently in \cite{Cederwall}. Following these lines it would be nice to extend the construction of black/worm holes \cite{Palatini1}-\cite{Palatini2} but considering T-duality invariant approaches as in \cite{BlairArv}.  

\item The results of this work can be extended to the next leading order in fermions. It is well known that when 4-fermion contributions are considered in the RS Lagrangian, the torsion requieres a specific fermionic contribution \cite{Tanii}. It would be interesting to study if it is possible to replace these contributions with fermionic extra terms deforming the non-metricity tensor.
\end{itemize}

\subsection*{Acknowledgements}
C.B acknowledges CONICET for support. The work of E.L is supported by the Croatian Science Foundation project IP-2019-04-4168. E.L also thanks CONICET for the support during the initial stages of this project.

\appendix
\section{Conventions}
\label{Conventions}
We use Greek letters as space-time indices, $\mu=0,\dots,3$, and Latin letters as flat indices, $a=0,\dots,3$. We use the notation $T_{[ab]}=\frac12( T_{ab} - T_{ab})$ and its generalization to further indices. 

The fundamental fields are a 4-dimensional vielbein, $e_{\mu}{}^{a}$ and a 4-dimensional Majorana gravitino, $\psi_{a}$ (when we write fermionic quantities, we do not explicitly write  spinorial indices). We use the notation: $\gamma^{ab}  =  \gamma^{[a} \gamma^{b]} = \frac12 \gamma^{a} \gamma^{b} - \frac12 \gamma^{b} \gamma^{a}$ and its generalization for more indices.

The covariant derivative is given by $\nabla_{\mu}$ and our convention is that $\nabla_{\mu}$ covariantizes the derivative of an object with respect to all the symmetries that the object transforms under.

\section{Gamma matrix relations}
\label{Gamma}
The Clifford algebra (\ref{Calgebra}) determines the following identities for the gamma matrices,
 \begin{subequations}\label{iden}
\begin{align}
\gamma_{{ab}} \gamma_{c} &=  \gamma_{{ab c}} + 2 \gamma_{[a} \eta_{b]c} \; ,\\
\gamma_{a}\gamma_{{bc}}&=\gamma_{{abc}}+2\eta_{a[b}\gamma_{c]}\; , \\ 
  \gamma_{{ab}} \gamma^{{c d}} & =  \gamma_{{ab}}{}^{{c d}} + 4 \gamma_{[a}{}^{[d} \eta_{b]}{}^{c]} 
+ 2 \eta_{[b}{}^{[c}\, \eta_{a]}{}^{d]} \; ,\\
 \gamma_{{ab}} \gamma^{{c d e}} & = \gamma_{{ab}}{}^{{c d e}} +6 \gamma_{[a}{}^{[{d e}} \eta_{b]}{}^{c]}  
+ 6 \gamma^{[e} \eta_{[b}{}^{c}\, \eta_{a]}{}^{d]} \; ,\\
  \gamma_{{ab c}} \gamma^{{d e}} & =  \gamma_{{ab c}}{}^{{d e}} +6 \gamma_{[{ab}}{}^{[e} \eta_{c]}{}^{d]} 
+ 6 \gamma_{[a} \eta_{c}{}^{[d}\, \eta_{b]}{}^{e]} \; .
\end{align}
 \end{subequations}

\end{document}